\documentclass[aps,prl,twocolumn,superscriptaddress,showpacs,floatfix]{revtex4-2}
\usepackage{graphicx,bm,epsfig,textcomp,color,dcolumn,setspace,array,calrsfs,mathrsfs,amsmath,amssymb,mathptmx,gensymb,booktabs,url,bibentry,natbib,subfigure,braket,lipsum,xcolor,relsize,diagbox}
\usepackage[T1]{fontenc}
\usepackage[mathscr]{eucal}
\usepackage[bookmarks=true,colorlinks=true,allcolors=blue,plainpages=false,pdfpagelabels,final,breaklinks=true]{hyperref}
\usepackage{cleveref}
\usepackage{tikz}

\DeclareMathAlphabet{\mathcal}{OMS}{cmsy}{m}{n}
\makeatletter
\newcommand{\supplementarytableofcontents}{%
  \@starttoc{stoc}%
}
\def\fps@figure{!t}
\def\fps@table{!t}
\makeatother 

\begin{document}

\title{Non-Abelian Anyon Braiding with Quantum-Antidot Interferometry}

\author{Junyu Tang}
\affiliation{International Center for Quantum Materials, School of Physics, Peking University, 
Beijing 100871, China}

\author{Gang v.~Chen}
\email{chenxray@pku.edu.cn}
\affiliation{International Center for Quantum Materials, School of Physics, Peking University, 
Beijing 100871, China}
\affiliation{Beijing Key Laboratory of Quantum Devices, Peking University, Beijing 100871, China}
\affiliation{Collaborative Innovation Center of Quantum Matter, 100871, Beijing, China}

\begin{abstract}
Conventional Fabry--Perot interferometry accesses only full braids of anyons and 
therefore cannot directly probe the elementary \(\pi\)-rotation exchange. 
Motivated by the recent quantum-antidot proposal for the Abelian anyons, 
we propose an interferometry for probing the elementary exchanges 
of non-abelian anyons using two gate-controlled quantum antidots. 
By tuning two antidots independently, 
the device realizes distinct cooperative tunnelling processes, 
which correspond to different braids of non-abelian anyons. 
For the unresolved local fusion channels, the difference 
between the interference signals of the single and double cooperative processes 
allows us to measure the elementary exchange of the non-abelian anyons, 
providing a direct probe of their non-abelian statistics 
and topological spins. For the resolved local fusion channels, 
the double-cooperative process is further distinguished 
by a reduced interference amplitude. 
Our work provides a promising and practical route 
for manipulating and detecting non-abelian braiding 
with the fundamental fractional statistics.
\end{abstract}

\maketitle

\noindent\textit{Introduction.}---The observation of anyonic braiding 
in a ${\nu=1/3}$ fractional quantum Hall (FQH) interferometer~\cite{Nakamura2020} 
provided the direct evidence for fractional statistics. 
Two tunnelling paths from source to drain encircle the central FQH droplet 
and generate a phase shift set by the braiding between bulk anyons 
and edge tunnelling anyons. As emphasized in Ref.~\onlinecite{Read2024}, 
however, the measured phase shift ${2\theta=2\pi/3}$ corresponds to a full braid, 
(i.e., two successive elementary exchanges) 
and is topologically equivalent to one anyon circling another by $2\pi$. 
Such a process does not isolate the elementary exchange phase. 
A full braid cannot even distinguish bosons from fermions since $(+1)^2=(-1)^2$. 
The more fundamental statistical phase, that corresponds to the elementary exchange 
or a $\pi$ rotation between anyons, is $\theta$, not $2\theta$.

Recently, Kivelson \textit{et al} proposed a modified interferometer 
that directly probes the elementary exchange of Abelian anyons~\cite{Kivelson_PRL_2025}. 
A quantum antidot (QAD) inserted at the midpoint of a quantum point contact (QPC) is 
tuned through resonance by a gate voltage, converting the direct tunnelling 
into a cooperative process. An anyon $A$ tunnels from one edge onto the QAD, 
while the other anyon $B$, initially bound to the QAD, exits to the opposite edge. 
After propagating around the interferometer bulk, $B$ returns to the initial position of $A$. 
Macroscopically, it corresponds to one effective anyon encircling the bulk once. 
Microscopically, however, the cooperative process contains an elementary exchange 
between $A$ and $B$, so the interference isolates a clean elementary exchange phase $\theta$.

\begin{figure}[!t] 
    \centering
  \includegraphics[width=1\linewidth]{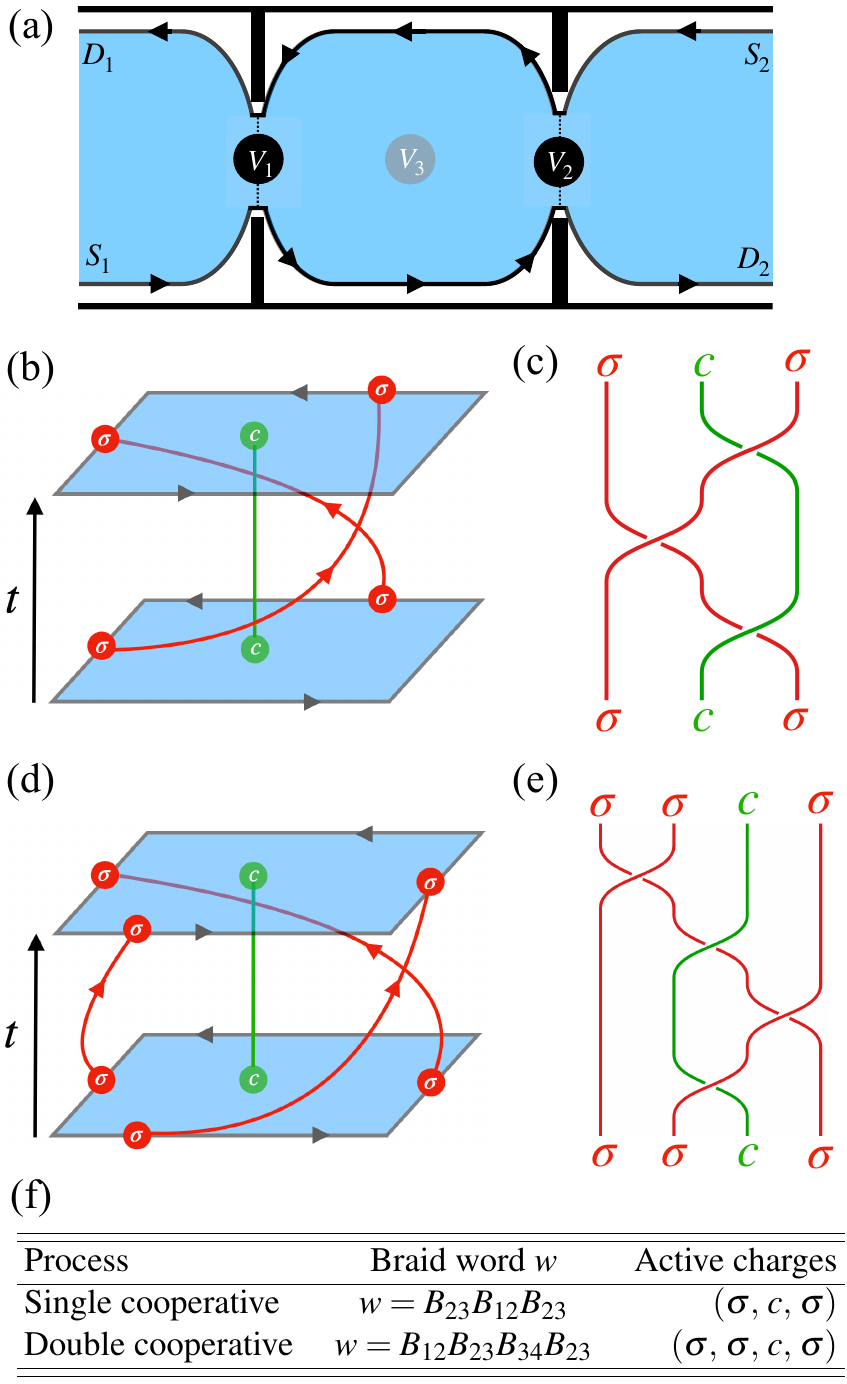}
  \caption{(a) The anyon interferometer. 
  The left and right QADs (black circles), controlled by $V_1$ and $V_2$, 
  determine the tunnelling processes at the two QPCs. 
  The middle QAD (gray circle), controlled by $V_3$, localizes the bulk anyon. 
  $S_{1/2}$ and $D_{1/2}$ denote the sources and drains. 
  The blue region is the FQH droplet. 
  (b,c) Single-cooperative process and its anyon world lines, 
  shown for the left (right) QPC being the cooperative (direct) tunnelling. 
  (d,e) Double-cooperative process and its anyon world lines, 
  with both QPCs tuned to cooperative tunnelling. (f) Summary of the braiding processes and the involved topological charges. Here $c$ denotes the total topological charge of the bulk quasiparticles, 
  and $\sigma$ the edge anyon.}
  \label{fig:interferometer}
\end{figure}

Here we extend this proposal to the non-abelian anyons 
by introducing two additional QADs~\cite{TwoQAD}. 
Previous Fabry--Perot proposals for non-abelian 
interferometry~\cite{SternHalperin2006,Shtengel_PRL_2006,Bonderson_PRL_2006,Bishara_PRB_2009,Stern_PRB_2010,Halperin_PRB_2011}, 
like their Abelian counterparts, access only full braids. 
By contrast, an elementary exchange acts non-trivially 
in the fusion space and therefore naturally calls 
for a multi-anyon interferometric setting. 
We consider a modified interferometer with two QADs 
placed at the left and right QPCs, controlled independently by
the gate voltages $V_1$ and $V_2$, 
together with an additional bulk QAD that controls the localized anyon 
in the interferometer interior [Fig.~\ref{fig:interferometer}(a)]. 
Depending on the gate configuration, the interference involves braids of up to four non-Abelian anyons. Most importantly, the comparison between the single- and double-cooperative processes gives direct access to the elementary exchange, or equivalently the topological twist, of the non-Abelian anyons, beyond the full-braid information available in conventional Fabry--Perot interferometry. In the resolved-fusion regime, the double-cooperative process is further characterized by a reduced interference amplitude. Our proposal provides an experimentally accessible route to probing both the elementary exchange and the local fusion structure of non-Abelian anyons.

\noindent\textit{Formalism}---The electrical conductance $G_{xx}$  
measured between the sources ($S_{1/2}$) and drains ($D_{1/2}$) is 
governed by the interference between the two QAD-assisted tunnelling paths, 
\begin{align}
G_{xx}\propto& \bigl|(t_1U_1+t_2U_2+\cdots)|\Psi\rangle\bigr|^2\notag\\
&=2|t|^2+2|t|^2\mathrm{Re}\!\left[e^{i\alpha}\langle\Psi|U_1^{-1}U_2|\Psi\rangle\right]
+\cdots,
\end{align}
where ${t_1=t_2=t}$ are the tunnelling amplitudes at the two QPCs, 
$U_1$ and $U_2$ are the corresponding unitary evolutions for an anyon 
taking the two different paths from $S_1$ to $D_1$ (or equivalently from $S_2$ to $D_2$), 
$|\Psi\rangle$ is the initial state, and $e^{i\alpha}$ collects the Aharonov--Bohm phase 
acquired upon encircling the bulk in a magnetic field. 
To the leading order, the essential interference factor is
\begin{equation}
\mathcal{I}=\langle\Psi|\mathcal{U}|\Psi\rangle,\qquad \mathcal{U}\equiv U_1^{-1}U_2 .
\end{equation}
The operator $\mathcal{U}$ therefore represents the braiding process where 
an anyon circulates once along the edge around the central FQH droplet,
i.e. the blue region in Fig.~\ref{fig:interferometer},
whose edge supports the anyonic excitations.

In this Letter, we focus on the Ising anyons associated with the ${\nu=5/2}$ 
fractional quantum Hall state~\cite{MooreRead1991,Nayak_2008}. Their fusion rules are
\begin{align}
\psi \times \psi = I,\quad 
\psi \times \sigma = \sigma,\quad 
\sigma \times \sigma = I + \psi,
\label{eq:nontrivial_fusion}
\end{align}
where $\psi$ denotes the fermion, 
$\sigma$ denotes the Ising anyon and $I$ the vacuum. 
The Ising anyons can be well described directly by topological quantum field theory (TQFT)~\cite{Kitaev_2006}. The relevant topological data are the fusion matrix $F$ and the braiding matrix $R$~\cite{Nayak_2008,Kitaev_2006}
\begin{align*}
  F^{\sigma\sigma\sigma}_{\sigma} & = \frac{1}{\sqrt{2}}\begin{pmatrix}
    1 & 1\\
    1 & -1
  \end{pmatrix},\quad  R^{\sigma\sigma}=
  \begin{pmatrix}
    e^{-i\pi/8} & 0\\ 
    0 & e^{3i\pi/8}
  \end{pmatrix},
\end{align*}
which are written in the basis of the two fusion channels of two $\sigma$ anyons [Eq.~\eqref{eq:nontrivial_fusion}]. The $R$ matrix shows that an elementary exchange 
of two $\sigma$ anyons yields the phase factors $R^{\sigma\sigma}_I=e^{-i\pi/8}$ 
and $R^{\sigma\sigma}_\psi=e^{3i\pi/8}$ for the fixed fusion channels $I$ and $\psi$, respectively. 
Since all other $F$ matrices, such as $F^{I\sigma\sigma}_\psi$ and $F^{\psi\sigma\sigma}_\psi$, 
are trivial scalar, we denote the only nontrivial $F^{\sigma\sigma\sigma}_{\sigma}\equiv F$. 
It is straightforward to verify that $F=F^\dagger=F^{-1}$.

Tuning the QAD bound-state energy $\epsilon$ through
the resonance ($\epsilon>0\to\epsilon<0$) by the gate voltage $V_{1/2}$ 
converts the direct tunnelling (potential barrier regime) at a QPC 
into the cooperative tunnelling (potential well regime)~\cite{Kivelson_PRL_2025}, 
where the QADs possess the bound state of anyons. 
We first consider the case in which only one QPC is tuned into 
the cooperative regime [Fig.~\ref{fig:interferometer}(b)]. 
For example, let the left QPC be cooperative while the right QPC 
remains in the direct-tunnelling regime. 
Then circulating around the bulk once involves three relevant quasiparticles. 
When the edge anyon $\sigma_{\rm edge}$ reaches the left QAD, 
the anyon $\sigma_L$ initially bound at the left QAD leaves the antidot and 
propagates toward the right QPC; after direct tunnelling there, 
it returns to the original position of $\sigma_{\rm edge}$. 
The circulation process is thus effectively $\sigma_{\rm edge}\to\sigma_L\to\sigma_{\rm edge}$, 
while the bulk quasiparticles remain localized and are characterized by a total topological charge $c$. 
The corresponding world lines are shown in Fig.~\ref{fig:interferometer}(c). 
If the bulk anyons fuse to the vacuum $I$, 
the bulk world line is topologically trivial and may be omitted, 
leading to an effective braiding of two anyons only. 
If instead the right QPC is tuned to the cooperative regime 
while the left QPC remains direct, 
one can easily check that the resulting braid is equivalent to Fig.~\ref{fig:interferometer}(c) 
by the Yang--Baxter relation. 
We therefore expect the same interference signal 
whenever only one QPC is tuned to be cooperative tunneling, 
independent of which QPC is tuned to the cooperative regime. This provides an internal consistency check for the interferometer in the experimental setting.
On the other hand, 
when both QPCs are tuned to the cooperative-tunnelling regime [Fig.~\ref{fig:interferometer}(d)], 
four quasiparticles participate in the braiding process. 
The left and right QADs host the initially bound anyons $\sigma_L$ and $\sigma_R$, respectively, and the interference path generated by $\mathcal{U}$ corresponds to the circulation process $\sigma_{\rm edge} \to \sigma_R \to \sigma_L \to \sigma_{\rm edge}$ with the bulk charge $c$ remaining localized. 
The corresponding world lines are shown in Fig.~\ref{fig:interferometer}(e).


The world lines in Fig.~\ref{fig:interferometer} involve only elementary 
exchanges of the neighbouring quasiparticles, which we denote by $B_{ij}$ 
for the counter-clockwise exchange of the quasiparticles at the positions $i$ and $j$. 
The subscript $i$ and $j$ label the positions in the world-line diagram counted from left to right; 
they are tied to the position instead of the quasiparticles. 
The world line in the single cooperative process of Fig.~\ref{fig:interferometer}(c)
is represented by $B_{23}B_{12}B_{23}$ and is equivalent to $B_{12}B_{23}B_{12}$ 
by the Yang--Baxter relation. In our convention, the 
operations on the right act earlier in time.  
Accordingly, the world line in the double cooperative process 
of Fig.~\ref{fig:interferometer}(e) 
is represented by $B_{12}B_{23}B_{34}B_{23}$. The braid words of these two processes are
summarized at the bottom of Fig.~\ref{fig:interferometer}.


\noindent\textit{Case I: unresolved fusion channel.}---In this regime, 
we assume that the fusion outcome between the propagating anyons 
and the anyons bound to the QAD remains unresolved. 
Since the endpoint labels of the braid world lines  
match the initial labels, Figs.~\ref{fig:interferometer}(c) and (e) 
define the endomorphisms within a fixed topological-charge sector, and
the interferometric amplitude ${\mathcal{I}=\braket{\Psi|\mathcal{U}|\Psi}}$ 
can be evaluated as the categorical (diagrammatic) trace $\rm Tr_{cat}$ 
of the corresponding braid operator, represented by the braid closure of the world-line diagram. 
Here, ``braid closure'' is used in the standard knot-theoretic sense.
Each top endpoint is joined to the corresponding bottom endpoint by 
the boundary-parallel arcs outside the braid manifold~\cite{Reshetikhin1990,Shtengel_PRL_2006}.



\begin{figure}[!t]
  \centering
  \includegraphics[width=1\linewidth]{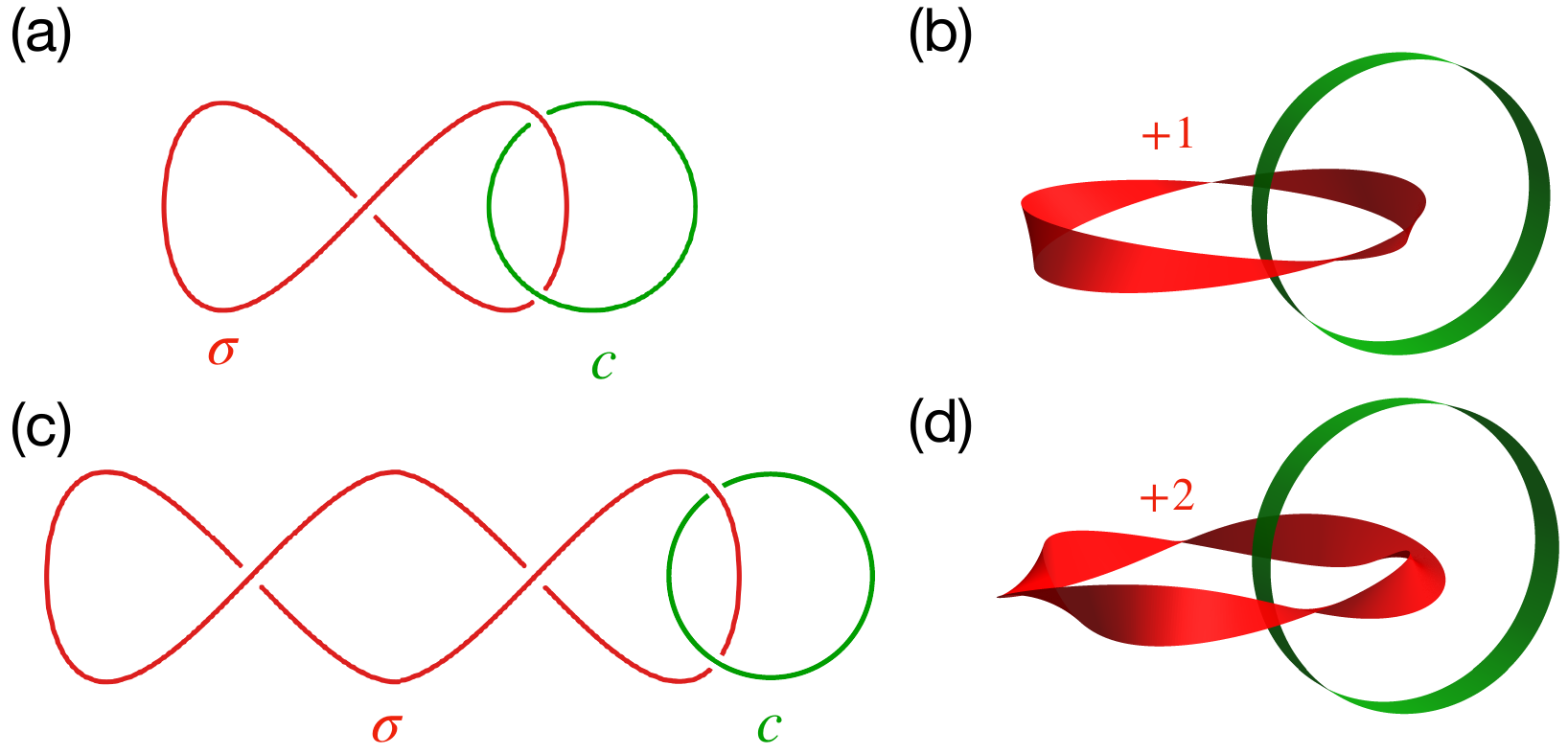}
  \caption{
(a) The blackboard framed diagram of the braid-world-line closure for the single-cooperative process, 
which contains a self crossing of Ising anyon $\sigma$ (red) with an additional Hopf link 
between the bulk charge $c$ (green) and $\sigma$. In ribbon language of (b), 
it is equivalent to one Hopf link with $\sigma$ carrying one additional positive twist (framing $+1$). 
(c) The blackboard framed diagram of the braid-world-line closure for the double-cooperative process, 
which contains two self-crossings and one Hopf link. Its ribbon representation in (d)
corresponds to the same Hopf link but with double twist (framing $+2$) in $\sigma$ ribbon.}
\label{fig:closure}
\end{figure}

For the single-cooperative process, 
the braid closure of the world line in Fig.~\ref{fig:interferometer}(c) gives 
Fig.~\ref{fig:closure}(a), which contains one self-crossing of the Ising anyon $\sigma$ 
and one Hopf link between the bulk charge $c$ and $\sigma$. 
For the double-cooperative process, the corresponding closure of Fig.~\ref{fig:interferometer}(e) is less straightforward, but after applying the Yang--Baxter relation to the lower-right part of the diagram and performing an isotopy deformation, one arrives at Fig.~\ref{fig:closure}(c). 
Relative to the single-cooperative case, the double-cooperative process contains one additional self-crossing of $\sigma$ with the same chirality, so the two crossings cannot be removed by a simple Reidemeister move.



A proper diagrammatic interpretation requires the framed world lines, or equivalently 
the ribbon representation, where a self-crossing in the blackboard framing is equivalent 
to a ribbon twist. The ribbon representations are shown in Fig.~\ref{fig:closure}(b) 
and Fig.~\ref{fig:closure}(d). 
Accordingly, Fig.~\ref{fig:closure}(a) can be interpreted as a Hopf link 
between the bulk charge $c$ and a $\sigma$ ribbon with framing $+1$, 
whereas Fig.~\ref{fig:closure}(c) corresponds to the same Hopf link 
with framing $+2$ on the $\sigma$ ribbon. 
A closed ribbon with framing $+|n|$ ($n\in \mathbb{Z}$) 
is obtained by rotating one end by $2\pi |n|$ before 
gluing the two ends together. 
This is different from a M\"obius strip that involves a rotation 
by odd times of $\pi$, namely $(2n+1)\pi$. 
Diagrammatically, a positively twisted $\sigma$ ribbon 
contributes the topological twist factor $\theta_\sigma$, 
while the Hopf link between $\sigma$ and $c$ contributes 
the monodromy factor $M_{\sigma c}$. 
We therefore obtain the interferometric amplitude 
for the corresponding braid word $w$ as 
\begin{align}
  \mathcal{I} = \frac{{\rm Tr_{cat}}[w]}{d_\sigma d_{c}} 
  =  \frac{{\rm Tr_{cat}}[\mathcal{S}]}{d_\sigma^2 d_{c}} \times {\rm Hopf}[\sigma,c] 
  = \theta_\sigma^n  M_{\sigma c},
  \label{eq:I_tm}
\end{align}
Here, $d_\sigma$ and $d_c$ are the quantum dimensions of $\sigma$ and $c$, 
required by the isotopy normalization. 
Each closed loop of topological charge contributes a normalization factor of 
quantum dimension~\cite{Simon_2023}. The additional factor of $d_\sigma$ in the third expression of Eq.~\eqref{eq:I_tm} 
arises from double counting the closed $\sigma$ loop 
after separating the Hopf-link and self-crossing contributions. 
$\mathcal{S}$ represents the one-self-crossing structure (twist) 
for the single-cooperative process ($n=1$) or the two-self-crossing structure 
(double twist) for the double-cooperative process ($n=2$). 
The topological twist factor $\theta_\sigma$ 
is related to the one self-crossing structure $\mathcal{S}$ (obtained from one elementary exchange) 
as $\theta_\sigma  = {\rm Tr_{cat}}[\mathcal{S}]/d_\sigma={\rm tr}[R^{\sigma\sigma}]/\sqrt{2}$. 
The Hopf link is related to the monodromy $M$ matrix through the ribbon identity as 
${\rm Hopf}[a,b]=\sum_x d_x R^{ab}_x R^{ba}_x=d_a d_b M_{ab}$, 
where fusion multiplicities have been omitted 
because they all equal to one for the physical non-abelian anyons. 
Using Eq.~\eqref{eq:I_tm}, we calculate the interferometric amplitudes 
for all the possible cases and the results are listed in Table~\ref{tab:interference}.

\begin{table}[t]
  \centering
  \begin{tabular}{c|c|c|c}
  \hline
  \hline
  \diagbox[width=9em]{Process}{Bulk charge} & $\quad c=I \quad $ & $\quad c=\psi\quad $ & $\quad c=\sigma\quad $   \\
  \hline
  Single cooperative & $e^{i\pi/8}$ & $-e^{i\pi/8} $ & $0$\\
  \hline
  Double cooperative & $e^{i\pi/4}$ & $-e^{i\pi/4}$ & $0$\\
  \hline  
  \hline
\end{tabular}
\caption{Interference amplitude $\mathcal{I}$ for each possible bulk topological charge 
$c \in \{ I,\psi,\sigma\}$ of single cooperative and double cooperative processes.}
\label{tab:interference}
\end{table}


For the Ising anyons, the elementary exchange gives a twist factor ${\theta_\sigma=e^{i\pi/8}}$, 
which corresponds to $1/16$ topological spin according to the spin-statistics 
theorem~\cite{Nayak_2008,Kitaev_2006}. 
The symmetric monodromy $M_{\sigma c}$ can be calculated from the $R$ matrix 
and the fusion rules, giving ${M_{\sigma I}=1}$, ${M_{\sigma\psi}=-1}$, 
and ${M_{\sigma\sigma}=0}$. 
Therefore, if the bulk charge fuses to $I$, namely ${c=I}$, 
the elementary exchange can be directly measured from the interferometric amplitude 
${\mathcal{I}=\theta_\sigma=e^{i\pi/8}}$ in the single-cooperative regime. 
If the bulk charge fuses to $\sigma$, namely $c=\sigma$, 
the interference signal vanishes for both the single- and double-cooperative processes 
[last column of Table~\ref{tab:interference}] due to $M_{\sigma\sigma}=0$. 
The vanishing of $M_{\sigma\sigma}$ reflects destructive interference between 
the two unresolved fusion channels of two $\sigma$ anyons. 
If instead $c$ fuses to $\psi$, the interference acquires an additional minus sign, 
or equivalently a $\pm\pi$ phase shift due to ${M_{\sigma\psi}=-1}$ 
[second column of Table~\ref{tab:interference}]. 
It is worthwhile to point out that in the conventional Fabry--Perot interferometry 
without QADs in the QPCs~\cite{Shtengel_PRL_2006,Bonderson_PRL_2006}, 
for two Ising anyons one can only measure their full braids, 
yielding a null result because ${M_{\sigma\sigma}=0}$. 
Recent experiments suggest that non-abelian anyons may preserve a fixed fusion channel 
over a time scale of order $10\,{\rm ms}$~\cite{Aghaee2025,vanLoo2026SingleShotKitaev}; 
in principle, the extra minus sign associated with switching between the $\psi$ and $I$ 
fusion channels for bulk $c$ could therefore be resolved through fluctuations of 
the interference signal on the macroscopic time scales. 
As a sanity check, we further verify the results in Table~\ref{tab:interference} 
with the calculation from the matrix representation~\cite{SM}.


In the double-cooperative process the two elementary exchanges 
occur between different $\sigma$ anyons 
and are therefore not equivalent to a full braid; 
in particular, ${\theta_\sigma^2 \neq M_{\sigma\sigma}=0}$. 
Thus, both the single- and double-cooperative processes can 
access the elementary exchange phase $\theta_\sigma$ with different exponents $n$ [Eq.~\eqref{eq:I_tm}]. 
This allows us to extract the twist factor from the phase difference between the single- and double-cooperative processes
\begin{align}
  \frac{\mathcal{I}^{\rm double}}{\mathcal{I}^{\rm single}}=\theta_\sigma,\quad c=I,\psi \label{eq:extraction}
\end{align}
by tuning only the gate voltages $V_1$ and $V_2$, 
while keeping all the other parameters fixed (such as the magnetic field and the bulk-QAD gate voltage $V_3$). The phase difference $\arg \theta_\sigma$ then yields the elementary exchange phase between the two Ising anyons in the unresolved-fusion regime. Such a protocol therefore provides a robust route to measuring the elementary exchange and topological spin of the non-abelian anyons~\cite{effect_of_V}.


\noindent\textit{Case II: resolved fusion channel}---We next consider the other regime, 
where the local fusion channel between a QAD-bound anyon 
and the propagating anyon is resolved and fixed. Physically, 
when two $\sigma$ anyons are brought sufficiently close near a QAD, 
the direct wave-function overlap~\cite{Gharavi2016, Baraban2009} 
and short-range interactions that mediate the tunnelling of 
topological charge between them~\cite{Bonderson2009,Cheng2009} 
can lift the degeneracy between the fusion channels $I$ and $\psi$. 
Then, the interferometer no longer probes a trace over the unresolved local fusion sectors, 
but rather the braiding evolution of a definite internal state 
in the fixed local fusion channel $x\in\{I,\psi\}$~\cite{fusion}. 
We now assume that, whenever a cooperative tunnelling event occurs, 
the local fusion channel between the propagating anyon 
and the corresponding QAD-bound anyon is resolved and fixed. 
The fusion channel between the central bulk charge $c$ 
and the edge anyon $\sigma$ still remains 
unresolved because they stay well separated during
the edge propagation. 
Therefore, for the single cooperative process, 
the monodromy factor $M_{\sigma c}$ in Eq.~\eqref{eq:I_tm} is unchanged, 
whereas the local twist factor $\theta_\sigma$ is replaced 
by the corresponding fixed-channel braiding eigenvalue $R^{\sigma\sigma}_x$.


As for the double-cooperative process, 
there are again only two resolved sectors rather than four independent ones, 
because the effective braiding acts in the fusion space of three $\sigma$ anyons 
at the edge or QADs, whose total topological charge is fixed to $\sigma$ 
by the Ising fusion rules. 
The relevant Hilbert space therefore remains two-dimensional, 
labelled by the intermediate fusion channel $x\in\{I,\psi\}$ 
of one adjacent $\sigma\times\sigma$ pair. 
Importantly, the second elementary exchange acts on a different neighbouring pair, 
so one cannot simply replace $\theta_\sigma^2$ by $(R_x^{\sigma\sigma})^2$. 
Instead, one must evaluate the diagonal matrix element of the two-exchange operator 
in the three-anyon fusion space. 
In the basis where the first two $\sigma$ anyons fuse to $x\in\{I,\psi\}$, 
we have $B_{12}=R$ and $B_{23}=FRF^{-1}$. 
The interference amplitudes in the resolved-fusion regime are therefore~\cite{order}
\begin{subequations}
\begin{align}
\mathcal I_x^{\rm single} &= R^{\sigma\sigma}_x\, M_{\sigma c},\\
\mathcal I_x^{\rm double} &= \bigl[R^{\sigma\sigma} (FR^{\sigma\sigma}F^{-1}) \bigr]_{xx}\, M_{\sigma c},\label{eq:I_double_resolved}
\end{align}
\end{subequations}
where the fixed intermediate fusion channel is $x\in\{I,\psi\}$. We calculate the interference amplitudes for all the possible intermediate fusion channels and bulk charges. The results are summarized in Table~\ref{tab:fusion_resolved}.

\begin{table}[!t]
  \centering
  \begin{tabular}{c|c|c|c}
  \hline
  \hline
  \diagbox[width=10em]{Fusion channel}{Bulk charge} & $\quad c=I \quad $ & $\quad c=\psi\quad $ & $\quad c=\sigma\quad $    \\
  \hline
  $x=I$ (single) & $e^{-i\pi /8}$ & $-e^{-i\pi/8} $ & $0$\\
  \hline
  $x=\psi$ (single) & $e^{3i\pi /8}$ & $-e^{3i\pi/8}$ & $0$\\
  \hline
  $x=I$ (double) & $\frac{1}{\sqrt 2}$ & $ -\frac{1}{\sqrt 2} $ & $0$\\
  \hline
  $x=\psi$ (double) & $\frac{i}{\sqrt 2}$ & $-\frac{i}{\sqrt 2}$ & $0$\\
  \hline  
  \hline
\end{tabular}
\caption{Interference amplitude $\mathcal{I}$ for each possible bulk topological charge 
$c \in \{ I,\psi,\sigma\}$ and fusion channel $x \in \{I,\psi\}$ of single and double cooperative processes.}
\label{tab:fusion_resolved}
\end{table}


In the resolved-fusion regime, the double-cooperative interference 
signal $\mathcal{I}$ is no longer a pure phase of unit modulus 
since the second exchange rotates the selected local fusion state 
into a coherent superposition within the same three-anyon fusion space, 
while the interferometer measures only its overlap with the initial state. 
This reduces the interference amplitude by a factor of $1/\sqrt{2}$ 
[last two rows of Table~\ref{tab:fusion_resolved}]. In this sense, 
the double-cooperative process in the resolved-fusion regime is 
distinguished from all other cases not only by its phase shift, 
but also by its reduced amplitude. 
By contrast, the single-cooperative interference signal 
remains a pure phase factor with unit modulus 
(except for the ${c=\sigma}$ case, where interference vanishes). 
The double-cooperative process is therefore more sensitive to 
the local fusion channel than the single-cooperative one. Additionally, from Table~\ref{tab:fusion_resolved}, we can find that when the interference signal is nonvanishing $(c\neq \sigma)$, the interference amplitude difference is characterized by a universal ratio
\begin{align}
  \frac{\mathcal{I}^{\rm double}_x}{\mathcal{I}^{\rm single}_x}=\frac{e^{i\pi/8}}{\sqrt{2}}=\frac{\theta_\sigma}{\sqrt{2}}\quad c=I,\psi,
\end{align}
which is independent of the intermediate fusion channel $x\in \{I,\psi\}$ and bulk charge $c$. Comparing with the Eq.~\eqref{eq:extraction} in the unresolved-fusion regime, the interference amplitude difference in the resolved-fusion regime has an additional reduction factor of $1/\sqrt{2}$, providing a clear characteristic for distinguishing the resolved-fusion regime from the unresolved-fusion regime.



\noindent\textit{Discussion.}---For the double-cooperative processes, Eq.~\eqref{eq:I_double_resolved} requires 
a coherent description for the three involved Ising anyons, which is natural for non-abelian anyons. 
Once a fusion channel is selected, the corresponding topological charge is stored nonlocally 
and is therefore protected against local perturbations. 
If this nonlocal fusion coherence is, however, 
lost or broken during the edge propagation, 
the double-cooperative process reduces to a sequentially resolved limit~\cite{decoherence}, 
where the two resolved fusion should be treated independently. 
In the resolved fusion regime, such decoherence effect makes the interference amplitude 
distinct from Eq.~\eqref{eq:I_double_resolved}, where $\mathcal{I}$ becomes a matrix [Eq.S37]
arising from the two independent exchanges at the two QADs [Table S2]. 
On the contrary, in the unresolved fusion regime, the decoherence effect 
does not change the interference amplitude given by Eq.~\eqref{eq:I_tm}. 
We leave the detailed analysis of the decoherence effects in the SM.

So far we have focused on the neutral Ising sector. In the physical ${\nu=5/2}$ Moore--Read state, the $\sigma$ quasiparticle also carries an Abelian $U(1)$ charge-flux sector with charge $e/4$ and flux $\Phi_0/2$~\cite{Bonderson_PRL_2006,Shtengel_PRL_2006}, so each elementary exchange acquires an additional phase $\pi/8$. The resulting phase shifts owing to the $U(1)$ sector are summarized in Table~\ref{tab:additional_phase}. This Abelian contribution changes the absolute interference phase but does not affect our extraction protocol: comparing the single- and double-cooperative processes still isolates the elementary exchange of the physical Ising anyons. Since the minimal-charge $e/4$ sector has the lowest scaling dimension, it is expected to dominate the tunnelling~\cite{Bonderson2007,Fendley_PRB_2007,Veillon2024}, thus the higher-flux sector $(\sigma, 3e/4,3\Phi_0/2)$ does not concern us.

\begin{table}[!t]
  \centering
  \begin{tabular}{c|c|c}
  \hline
  \hline
  \diagbox[width=9em]{Process}{Bulk charge} & $\quad c=I,\psi \quad $ & $\quad c=\sigma\quad $   \\
  \hline
  Single cooperative & $e^{i(4m + 1) \pi/8 }$ & $e^{i(4m + 3) \pi/8 } $\\
  \hline
  Double cooperative & $e^{i(4m + 2) \pi/8 }$ & $e^{i(4m + 4) \pi/8 }$ \\
  \hline  
  \hline
\end{tabular}
\caption{Additional $U(1)$ phase factors $(m\in \mathbb{Z})$ for Ising anyons in the $\nu=5/2$ FQH state with different bulk charges. The results are obtained by counting the number of exchanges between the $\sigma$ anyons and the bulk charge $c$, as well as between the $\sigma$ anyons themselves. For $c=I$ or $c=\psi$, we may regard $c$ as fused from even number $2m$ of $\sigma$ anyons, whereas for $c=\sigma$, it can be regarded as fused from odd number $2m+1$ of $\sigma$ anyons.}
\label{tab:additional_phase}
\end{table}

In summary, we have proposed a two-QAD interferometer for probing the non-abelian anyon braiding. Independent control of the two QADs enables single- and double-cooperative tunnelling processes, whose interference signals encode the elementary exchange through the twist factor. Although our analysis has focused on Ising anyons relevant to the $\nu=5/2$ Moore--Read state, the same interferometric principle should be extendable to other non-abelian anyons. A particularly interesting direction is Fibonacci-type anyons, which is related to the $\nu=12/5$ FQH state~\cite{ReadRezayi_PRB_1999,Bonderson_PRL_2006,Mong_PRB_2017,Zhu_PRL_2015}.

\noindent\textit{Acknowledgments}--We thank Hong-hao Song and Jiahao Yang 
for the fruitful discussions. This work is supported by 
 Quantum Science and Technology-National Science and Technology 
 Major Project (grant No.~2025ZD0300500) 
and NSFC with Grants No.~92565110 and No.~12574061, 
BJNSF with No.~F261004.

\bibliographystyle{apsrev4-2}
\nocite{apsrev42Control}
\bibliography{references.bib}

\clearpage
\onecolumngrid
\setcounter{page}{1}
\renewcommand{\thepage}{S\arabic{page}}
\setcounter{section}{0}
\renewcommand{\thesection}{S\arabic{section}}
\setcounter{equation}{0}
\renewcommand{\theequation}{S\arabic{equation}}
\setcounter{figure}{0}
\renewcommand{\thefigure}{S\arabic{figure}}
\setcounter{table}{0}
\renewcommand{\thetable}{S\arabic{table}}
\begin{center}
{\large\bf Supplementary Material for}\\[0.5em]
{\large\bf  Non-Abelian Anyon Braiding
with Quantum-Antidot Interferometry}\\[1.0em]

Junyu Tang$^{1}$ and Gang v. Chen$^{1,2,3}$\\[0.5em]

{\small
$^{1}$\textit{International Center for Quantum Materials, School of Physics,
Peking University, Beijing 100871, China}\\
$^{2}$\textit{Beijing Key Laboratory of Quantum Devices, Peking University, Beijing 100871, China}\\
$^{3}$\textit{Collaborative Innovation Center of Quantum Matter, 100871, Beijing, China}
}
\end{center}
\tableofcontents

\vspace{2cm}

\twocolumngrid

In the supplementary material, we present a matrix-based verification of the results obtained in the main text. We also provide a detailed analysis of decoherence effects in the double-cooperative process. In the resolved-fusion regime, decoherence leads to different interference amplitudes, whereas in the unresolved-fusion regime the interference amplitude remains unchanged. Finally, we discuss the possibility of generalizing the interferometry to other platforms hosting non-Abelian anyons, such as topological superconductors and chiral spin liquids.

\section{Matrix-based approach}
We emphasize that the results obtained in the unresolved regime [Table~II] can also be obtained by taking the trace of the matrices representing the corresponding braiding operators in the fusion basis. The results agree well with our diagrammatic calculus. The diagrammatic approach presented in the main text provides a more intuitive and direct way to understand the interference patterns. This approach also makes it easier to understand their topological nature, as well as the role of framing and ribbon structures in the world-line diagrams, whereas the conventional matrix-based approach requires more careful bookkeeping of the fusion basis and the corresponding $F$-moves. Nevertheless, we present the matrix-based verification below for completeness and as a sanity check of our diagrammatic approach.

Before presenting the matrix-based verification, we first clarify an important
relation between the categorical trace used in the diagrammatic calculation
and the ordinary matrix trace used in a chosen fusion basis. For a braid
operator $W$ acting in the fusion space of a set of anyons with possible
total topological charge $q$, we denote the corresponding fixed-total-charge
fusion space by $V^q$. In a fusion-tree basis, $W$ is represented by an
ordinary matrix $W_q$ on $V^q$. The categorical trace, however, is obtained
by closing not only the internal fusion lines but also the external total
charge line. Therefore each fixed-$q$ block is weighted by the quantum
dimension $d_q$. Following the standard anyon diagrammatic formalism,
the categorical (quantum) trace of an operator decomposed into fixed total-charge sectors
is related to the ordinary matrix trace by~\cite{Bonderson2007}
\begin{align}
  {\rm Tr}_{\rm cat}(W)
  =
  \sum_q d_q\,{\rm tr}_{V^q}(W_q),
  \label{eq:cat_trace_vs_ord_trace}
\end{align}
where ${\rm tr}_{V^q}$ denotes the ordinary matrix trace in the fusion space
$V^q$. For Ising anyon braiding considered here (six cases, see Table~\ref{tab:cat_vs_ord_trace}), the total charge $q$ can either be fixed to a single $\sigma$ or $I\oplus \psi$. For Ising anyons, we have $d_I=d_\psi=1$ and $d_\sigma=\sqrt{2}$. Consequently, if the total charge is fixed to $\sigma$, we have the following relation
\begin{align}
  {\rm Tr}_{\rm cat}(W_\sigma)
  =
  d_\sigma\,{\rm tr}_{V^\sigma}(W_\sigma) = \sqrt{2}\, {\rm tr}(W_\sigma) .
\end{align}
On the other hand, when the total charge sectors is
$I\oplus \psi$, the categorical trace coincides with the ordinary trace as
\begin{align}
  {\rm Tr}_{\rm cat}(W)
  =
  d_I\,{\rm tr}_{V^I}(W_I) + d_\psi\,{\rm tr}_{V^\psi}(W_\psi) = {\rm tr}(W).
\end{align}
Note that even the total charge $q$ has been fixed, the braid word in the subspaces $W_I$, $W_\psi$ and $W_\sigma$ can still be a matrix as the internal fusion space could be multi-dimensional. The total braid word $W$ is then a block-diagonal matrix with blocks $W_I$, $W_\psi$ and $W_\sigma$, namely $W=\sum_q \oplus W_q$. To summarize, the ordinary trace corresponds to summing over the diagonal elements of the matrix, while the categorical trace is obtained by closing the world lines and applying the isotopy normalization, which in the matrix language gives an additional factor of $d_q$ for each external total charge sector $q$.

One should not confuse this extra nontrivial $d_\sigma$ factor from closing the word line of external total charge $\sigma$ with the one in the diagrammatic normalization for calculating the interference amplitude
\begin{align}
  \mathcal I
  =
  \frac{{\rm Tr}_{\rm cat}(W)}{d_\sigma d_c},
  \label{eq:interference_normalization}
\end{align}
From the above expression, it's easy to see that if the total charge is fixed to $\sigma$, then after replacing the categorical trace by the ordinary trace, the $d_\sigma$ in the denominator will be canceled out. For the six cases considered in the main text, the relation between categorical and ordinary traces is summarized in Table~\ref{tab:cat_vs_ord_trace}.

\begin{table}[!t]
\centering
\begin{tabular}{c c c}
\hline\hline
Process & Total charge & Relation \\
\hline
Single, $c=I$
&
$I\oplus\psi$
&
${\rm Tr}_{\rm cat} ={\rm tr}$
\\
Single, $c=\psi$
&
$I\oplus\psi$
&
${\rm Tr}_{\rm cat} ={\rm tr} $
\\
Single, $c=\sigma$
&
$\sigma$
&
${\rm Tr}_{\rm cat} =d_\sigma\,{\rm tr}$
\\
Double, $c=I$
&
$\sigma$
&
${\rm Tr}_{\rm cat} =d_\sigma\,{\rm tr} $
\\
Double, $c=\psi$
&
$\sigma$
&
${\rm Tr}_{\rm cat} =d_\sigma\,{\rm tr} $
\\
Double, $c=\sigma$
&
$I\oplus\psi$
&
${\rm Tr}_{\rm cat} ={\rm tr} $
\\
\hline\hline
\end{tabular}
\caption{
Relation between the categorical trace ${\rm Tr}_{\rm cat}$ and the ordinary matrix trace ${\rm tr}_{\rm ord}$ for the
six unresolved-fusion cases.
}
\label{tab:cat_vs_ord_trace}
\end{table}

\section{Verification: Unresolved-fusion regime}

\subsection{Single-cooperative process}
In the single-cooperative case, when $c=I$, the braid world lines in Fig.~1(c) reduce to a single elementary exchange of two $\sigma$ anyons. In the fusion space of two Ising anyons, we have
\begin{align}
  \frac{{\rm tr}[R^{\sigma\sigma}]}{d_\sigma}=\theta_\sigma = e^{i\pi/8}
\end{align}
which directly gives the elementary exchange phase $\pi/8$ shown in Table~II.

When $c=\sigma$, in the fusion space of three Ising anyons, one can obtain the braid between any pair of anyons by a standard $F$ move. We denote the braid between the first two $\sigma$ anyons by $B_{12}=R$, and the braid between the second and third $\sigma$ anyons by $B_{23}=FRF^{-1}$. Here, the nontrivial $F$ matrix is denoted as $F=F^{\sigma\sigma\sigma}_\sigma$. The braid closure of Fig.~\ref{fig:interferometer}(c) corresponds to the braid word $w=B_{23}B_{12}B_{23}$, which is equivalent to $B_{12}B_{23}B_{12}$ by the Yang--Baxter relation. In either case, by using the cyclic property of the trace, one can verify that ${\rm Tr_{cat}}[w]/d_\sigma^2$ is equal to ${\rm tr_{cat}}[R(FRF)R]/d_\sigma$ (note that $F^{-1}=F$). In the matrix representation, the categorical trace corresponds to the quantum trace. Since all particles in the intermediate fusion channels ($\psi, I$) have quantum dimension equal to $1$, the quantum trace reduces to the ordinary trace. It is then easy to verify that
\begin{align}
  \frac{{\rm tr}[R(FRF)R]}{d_\sigma}=0
\end{align}
Thus, we indeed recover the null result for $c=\sigma$.

Last, we consider the case where $c=\psi$, which requires additional care because we are now dealing with different types of particles. In this case, the active 
topological charges are $(\sigma_1,\psi_2,\sigma_3)$. For convenience, we label the particles with numbers. One should not, however, confuse these labels with those in the braid word $w$. For example, in $w=B_{23}B_{12} B_{23}$, the first (rightmost) $B_{23}$ means the exchange between the particles in positions 2 and 3, namely $\psi_2$ and $\sigma_3$. But the next $B_{12}$ does not mean the exchange between $\sigma_1$ and $\psi_2$. Instead, it means the exchange between the particles in positions 1 and 2, which corresponds to the exchange between $\sigma_1$ and $\sigma_3$ (since $\sigma_3$ is now in position 2 after $B_{23}$). 

The relevant fusion 
space decomposes into two one-dimensional sectors labeled by the total 
charge $a\in\{I,\psi\}$. We take the fusion basis
\begin{align}
  \ket{a}=\ket{((\sigma_1\psi_2)_\sigma \sigma_3)_a},
\end{align}
where $\sigma$ and $\psi$ first fuse to $\sigma$, and then fuse with the other $\sigma$ to total charge $a$. The Hilbert space is thus denoted by $\ket{a}$. The needed Ising braiding data are $R^{\sigma\psi}_{\sigma}=R^{\psi\sigma}_{\sigma}=-i$. The corresponding $F$ moves involving a $\psi$ line are only one-dimensional
scalar transformations,
\begin{align}
  F^{\sigma\psi\sigma}_\psi=-F^{\sigma\psi\sigma}_I=-1,\quad  F^{\sigma\sigma\psi}_\psi= F^{\sigma\sigma\psi}_I=1,
\end{align}
Hence, they cancel in the combination $F^{-1}RF$. Equivalently, the result below is independent of these scalar $F$-symbol
conventions.

We now evaluate the braid word $w=B_{23}B_{12}B_{23}$. The first $B_{23}$ exchanges $\psi_2$ and $\sigma_3$ 
and changes the ordering from $(\sigma,\psi,\sigma)$ to 
$(\sigma,\sigma,\psi)$, contributing $R^{\psi\sigma}_{\sigma}$. In this 
intermediate ordering, $B_{12}$ exchanges the two $\sigma$ anyons. For a 
fixed total charge $a$, the fusion channel $x$ of these two $\sigma$ anyons 
is determined by the total charge $x\times\psi=a$, namely $x=a\times\psi$. Therefore this 
middle exchange contributes $R^{\sigma\sigma}_{a\times\psi}$. Finally, the 
leftmost $B_{23}$ exchanges $\sigma$ and $\psi$ and brings the ordering back 
to $(\sigma,\psi,\sigma)$, contributing $R^{\sigma\psi}_{\sigma}$. Thus the 
eigenvalue of $w$ in the sector $a$ is
\begin{align}
  \lambda_a
  =
  R^{\sigma\psi}_{\sigma}
  R^{\sigma\sigma}_{a\times\psi}
  R^{\psi\sigma}_{\sigma}
  =
  - R^{\sigma\sigma}_{a\times\psi}.
\end{align}
Explicitly,
\begin{align}
  \lambda_I=-R^{\sigma\sigma}_{\psi}=-e^{3i\pi/8},\qquad
  \lambda_{\psi}=-R^{\sigma\sigma}_{I}=-e^{-i\pi/8}. 
\end{align}
Since $d_I=d_\psi=1$, the quantum trace again reduces to the ordinary trace. 
Using $d_\sigma=\sqrt{2}$ and $d_\psi=1$, we obtain
\begin{align}
  \frac{{\rm tr}[B_{23}B_{12}B_{23}]}{d_\sigma d_\psi}
  =
  \frac{-e^{3i\pi/8}-e^{-i\pi/8}}{\sqrt{2}}
  =
  -e^{i\pi/8},
\end{align}
which agrees with the diagrammatic result
$\theta_\sigma M_{\sigma\psi}=-e^{i\pi/8}$ in Table~II.

\begin{figure}[!t]
    \centering
    \includegraphics[width=0.9\linewidth]{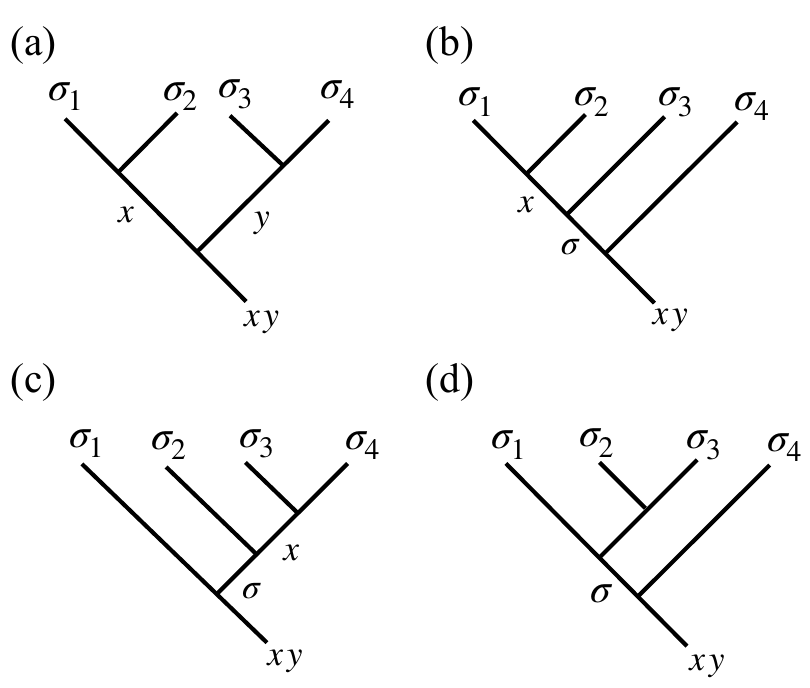}
    \caption{Four possible fusion trees for the four $\sigma$ anyons. Here $x,y \in \{I,\psi\}$ denote the intermediate fusion channels, and the final fusion channel is determined by $xy$.}
    \label{fig:fusion_tree}
\end{figure}

\subsection{Double-cooperative process}
The verification of the double-cooperative case involves four quasiparticles. We first consider the most interesting case where the bulk charge fuses to $c=\sigma$. To deal with four $\sigma$ anyons, we need to specify a fusion basis. The fusion space of four $\sigma$ anyons is four-dimensional and can be labeled by the intermediate fusion channels of two pairs of $\sigma$ anyons. We choose the fusion tree in Fig.~\ref{fig:fusion_tree}(a) as our starting basis, with intermediate fusion channels $x,y\in\{I,\psi\}$. Once $x$ and $y$ are specified, the total topological charge $xy$ is fixed by the fusion rules. This basis is particularly convenient because the exchanges of $(\sigma_1,\sigma_2)$ and $(\sigma_4,\sigma_3)$ act directly within the specified fusion channels. We denote the four-dimensional Hilbert space, spanned by the $xy$ sectors, by $\ket{xy=I}=\{\ket{II},\ket{\psi\psi}\}$ and $\ket{xy=\psi}=\{\ket{I\psi},\ket{\psi I}\}$. Accordingly, we find
$B_{12}={\rm diag}[R^{\sigma\sigma}_I,R^{\sigma\sigma}_\psi,R^{\sigma\sigma}_I,R^{\sigma\sigma}_\psi]$ and $B_{34}={\rm diag}[R^{\sigma\sigma}_I,R^{\sigma\sigma}_\psi,R^{\sigma\sigma}_\psi,R^{\sigma\sigma}_I]$. To obtain $B_{23}$, we transform to the basis in Fig.~\ref{fig:fusion_tree}(d), where $\sigma_2$ and $\sigma_3$ fuse first and then with $\sigma_1$. Note that the fusion outcome of three successive $\sigma$ anyons is fixed to be $\sigma$. To achieve this transformation, we apply $[F^{x\sigma\sigma}_{xy}]^{-1}$ to transition from Fig.~\ref{fig:fusion_tree}(a) to Fig.~\ref{fig:fusion_tree}(b), and apply $F^{\sigma\sigma\sigma}_\sigma\equiv F$ to transition from Fig.~\ref{fig:fusion_tree}(b) to Fig.~\ref{fig:fusion_tree}(d). By the standard procedure, i.e. applying $F^{-1}$ to transition back to the original basis, we find that the matrix for exchanging $\sigma_2$ and $\sigma_3$ is given by
\begin{align}
  B_{23}=FRF^{-1}\oplus FRF^{-1}.
\end{align}
The direct-sum structure reflects the fact that this basis transformation is independent of the final total topological charge $xy$, and therefore acts identically in the $xy=I$ and $xy=\psi$ sectors. As a sanity check, in the sector $x=y=\psi$, $B_{23}$ reproduces the familiar four-Majorana result: $FRF^{-1}$ rotates the state within the even-parity sector as $FRF^{-1}|\psi ,\psi \rangle =(e^{-\frac{3\pi }{8}i}|I,I\rangle +e^{\frac{\pi }{8}i}|\psi,\psi \rangle )/\sqrt{2}$, in agreement with (up to an overall phase convention) the standard Majorana-zero-mode representation of Ising anyons. 

Having obtained the matrix representations of the elementary braid operators, we can then verify that the quantum trace of the braid word $w=B_{12}B_{23}B_{34}B_{23}$, which corresponds to the double-cooperative process involving four $\sigma$ anyons, indeed reproduces the null results in Table~II, namely,
\begin{align}
  \frac{{\rm tr}[ B_{12} B_{23} B_{34} B_{23}]}{d^2_\sigma}=0.
\end{align}

When the central bulk charge fuses to $c=I$, the relevant fusion space reduces to two dimensions and we only need to deal with three Ising anyons, a case that has already been addressed above. In this situation, we can ignore the identity particle $I$, and the braid word in Fig.~\ref{fig:interferometer}(e) reduces to an effective braiding $w=B_{23}B_{12}$ between the three $\sigma$ anyons. As discussed previously, in the corresponding fusion space of three anyons, we have $B_{12}=R$ and $B_{23}=FRF^{-1}$. It is then easy to verify that
\begin{align}
  \frac{{\rm Tr}_{\rm cat}(w)}{d_\sigma}={\rm tr}[ B_{12} B_{23} ]=\theta_\sigma^2=e^{i\pi/4}.
\end{align}

Last, we consider the case where $c=\psi$ in the double-cooperative
process, which takes additional care. The active topological charges are now
$(\sigma_1,\sigma_2,\psi_3,\sigma_4)$, and the braid word is still $w=B_{12}B_{23}B_{34}B_{23}$.
Since $(\sigma\times\sigma)\times\psi\times\sigma=\sigma$, the total
topological charge is fixed to be $\sigma$, so the categorical trace is given by $d_\sigma$ times the ordinary trace in the fusion space of these four anyons. 

For a convenient discussion, we define
\begin{align}
  r_I=R^{\sigma\sigma}_I=e^{-i\pi/8},\qquad
  r_\psi=R^{\sigma\sigma}_\psi=e^{3i\pi/8},
\end{align}
and use
\begin{align}
  R^{\sigma\psi}_{\sigma}=R^{\psi\sigma}_{\sigma}=-i.
\end{align}
The only nontrivial two-dimensional $F$ move that appears is again
$F^{\sigma\sigma\sigma}_{\sigma}\equiv F$, so that
\begin{align}
  A\equiv FRF^{-1}=FRF
  =
  \frac{1}{2}
  \begin{pmatrix}
    r_I+r_\psi & r_I-r_\psi\\
    r_I-r_\psi & r_I+r_\psi
  \end{pmatrix}.
\end{align}

We now follow the braid word explicitly and keep track of the fusion labels
at each step. We start from the basis state
\begin{align}
  \ket{x}
  =
  \ket{(((\sigma_1\sigma_2)_x\psi_3)_{x\times\psi}\sigma_4)_\sigma},
  \qquad x\in\{I,\psi\}.
\end{align}
Here $x$ is the fusion channel of the first two $\sigma$ anyons. After
fusing with $\psi_3$, the total charge of the first three anyons becomes
$x\times\psi$. Thus, after $x$ fuses with $\psi$, the outcome $x\times \psi$ flips the label $I
\leftrightarrow \psi$ with respect to $x$.

The rightmost operator in $w$ is $B_{23}$. It exchanges $\sigma_2$ and
$\psi_3$, changing the ordering from
$(\sigma_1,\sigma_2,\psi_3,\sigma_4)$ to
$(\sigma_1,\psi_3,\sigma_2,\sigma_4)$. Strictly speaking, in the fusion
tree chosen above these two anyons do not fuse first. Therefore the action
of $B_{23}$ should be evaluated by first performing the appropriate
$F$ moves to a basis in which $\sigma_2$ and $\psi_3$ fuse first, then
applying the braiding eigenvalue $R^{\sigma\psi}_{\sigma}$, and finally
transforming back. However, since $\sigma\times\psi=\sigma$
contains only a single fusion channel, all the relevant $F$ moves are
one-dimensional scalar transformations. They therefore cancel between the
forward and inverse basis changes in the combination $F^{-1}RF=R$. Thus the
net effect of this exchange is simply the scalar factor
$R^{\sigma\psi}_{\sigma}$:
\begin{align}
  B_{23}\ket{x}
  =
  R^{\sigma\psi}_{\sigma}
  \ket{(((\sigma_1\psi_3)_\sigma\sigma_2)_{x\times\psi}\sigma_4)_\sigma}.
\end{align}
The important point is that the total charge of the first three anyons remains
unchanged by $B_{23}$. Therefore the corresponding intermediate label
in the effective three-$\sigma$ fusion space is
$x\times\psi$.

Next, the operator $B_{34}$ exchanges the two $\sigma$ anyons in positions
3 and 4 of the current ordering. Since $(\sigma_1\psi_3)$ fuses uniquely to
$\sigma$, we may regard $(\sigma_1\psi_3)_\sigma$, $\sigma_2$, and
$\sigma_4$ as three effective $\sigma$ anyons. In the basis where the first
two of these effective $\sigma$ anyons fuse first, the exchange of the
second and third effective $\sigma$ anyons is represented by
\begin{align}
  A\equiv FRF^{-1}.
\end{align}
Thus
\begin{align}
  B_{34}B_{23}\ket{x}
  =
  R^{\sigma\psi}_{\sigma}
  \sum_{y'=I,\psi}
  A_{y',\,x\times\psi}
  \ket{(((\sigma_1\psi_3)_\sigma\sigma_4)_{y'}\sigma_2)_\sigma}.
\end{align}
Here $y'$ labels the fusion channel of $(\sigma_1\psi_3)_\sigma$ and
$\sigma_4$ after the exchange $B_{34}$.

The next operator is again $B_{23}$. In the current ordering
$(\sigma_1,\psi_3,\sigma_4,\sigma_2)$, this exchanges $\psi_3$ and
$\sigma_4$. As before, these two anyons are not necessarily adjacent in the
chosen fusion tree, so the operation should be understood as an $F$ move to
the basis where $\psi_3$ and $\sigma_4$ fuse first, followed by
$R^{\psi\sigma}_{\sigma}$, and then the inverse $F$ move. Since
$\psi\times\sigma=\sigma$ has a unique fusion channel, the relevant
$F$ symbols are again scalar and cancel in $F^{-1}RF$. Hence this exchange
contributes only the scalar factor $R^{\psi\sigma}_{\sigma}$. After this exchange, the ordering becomes
$(\sigma_1,\sigma_4,\psi_3,\sigma_2)$. Suppose the fusion channel of the
first two $\sigma$ anyons in this final ordering is $u$. Then the total
charge of the first three anyons is $u\times\psi$. Since this total charge
must equal the label $y'$ before the exchange, we must set
\begin{align}
  y'=u\times\psi.
\end{align}
Therefore
\begin{align}
  B_{23}B_{34}B_{23}\ket{x}
  =&
  R^{\psi\sigma}_{\sigma}
  R^{\sigma\psi}_{\sigma}\times\notag\\
  &\sum_{u=I,\psi}
  A_{u\times\psi,\,x\times\psi}
  \ket{(((\sigma_1\sigma_4)_u\psi_3)_{u\times\psi}\sigma_2)_\sigma}.
\end{align}

Finally, the leftmost operator $B_{12}$ exchanges the first two $\sigma$
anyons in the final ordering. Previously, we have assumed their fusion channel is $u$, so this exchange
contributes $R^{\sigma\sigma}_u$. Hence the full braid operator has matrix
elements
\begin{align}
  W_{ux}
  =
  R^{\sigma\sigma}_{u}
  R^{\psi\sigma}_{\sigma}
  A_{u\times\psi,\,x\times\psi}
  R^{\sigma\psi}_{\sigma}=
  -R^{\sigma\sigma}_{u}
  A_{u\times\psi,\,x\times\psi}.
\end{align}

Therefore the ordinary trace over this two-dimensional fusion space is
\begin{align}
  {\rm tr}\,W =
  -\left[
  R^{\sigma\sigma}_I A_{\psi\psi}
  +
  R^{\sigma\sigma}_\psi A_{II}
  \right] =
  -\frac{1}{2}(r_I+r_\psi)^2.
\end{align}
Using
\begin{align}
  r_I+r_\psi
  =
  e^{-i\pi/8}+e^{3i\pi/8}
  =
  \sqrt{2}\,e^{i\pi/8},
\end{align}
we find
\begin{align}
  {\rm tr}\,W=-e^{i\pi/4}.
\end{align}
Because the total charge
of the fusion space is $\sigma$, restoring the categorical trace gives
${\rm Tr}_{\rm cat}[w]=d_\sigma\,{\rm tr}\,W$. Hence
\begin{align}
  \frac{{\rm Tr}_{\rm cat}[w]}{d_\sigma d_\psi}
  =
  \frac{d_\sigma\,{\rm tr}\,W}{d_\sigma}
  =
  -e^{i\pi/4},
\end{align}
where $d_\psi=1$. This agrees with the diagrammatic result $  \theta_\sigma^2 M_{\sigma\psi}=e^{i\pi/4}(-1)=-e^{i\pi/4}$
in Table~II.

\section{Verification: Resolved fusion regime}
We emphasize that the resolved-fusion calculation does not involve a
categorical trace over the full local fusion space. Instead, it computes a
normalized diagonal matrix element in a fixed fusion state. In this case, the normalized amplitude is given by
\begin{align}
  \frac{{\rm Tr}_{\rm cat}( P_x^{(q)} W)}
       {{\rm Tr}_{\rm cat}(P_x^{(q)})}
  =
  \frac{d_q\,{\rm tr}_{V^q}(P_x^{(q)} W)}
       {d_q\,{\rm tr}_{V^q}(P_x^{(q)})}
  ={\rm tr}_{V^q}(P_x^{(q)} W)
  =
  W^{(q)}_{xx}.
\end{align}
The projector onto the resolved local fusion channel $x$ in the
total-charge sector $q$ is
\begin{align}
  P_x^{(q)}
  =
  \ket{x;q}\bra{x;q}.
\end{align}
Here, $x$ labels the local fusion channel of the anyons involved in the braid, and $q$ is the total charge of these anyons. Note that $x$ may label either the intermediate fusion channel or the final fusion channel (such that $x=q$). The above projector expression is general for the resolved-fusion case; multiplicities have been omitted since they are all equal to one. It applies to both the single- and double-cooperative processes, as we discuss below. The extra quantum-dimension factor $d_q$ cancels in this case, so we only need the diagonal matrix element $W_{xx}^{(q)}$ of the braid word $W$ in the total-charge sector $q$.

\subsection{Single-cooperative process}
In the single-cooperative resolved case, the local resolved
state between the bound anyon and the edge anyon can be labeled simply by the final fusion outcome
\begin{align}
  \sigma\times\sigma \to x,\qquad x\in\{I,\psi\}.
\end{align}
In this case, the label $x$ is equivalent to the total charge $q$ of the two $\sigma$ anyons. The local braid (elementary exchange) acts diagonally in this fixed channel, contributing the amplitude
\begin{align}
  \mathcal A_x^{\rm single}
  =
  R^{\sigma\sigma}_x.
\end{align}
Note that the central bulk charge $c$ is still not locally fused with the edge and QAD-bound anyons and
therefore contributes the same unresolved monodromy factor $M_{\sigma c}$ as
in the main text. Hence, we have the interference amplitude
\begin{align}
  \mathcal I_x^{\rm single}
  =
  R^{\sigma\sigma}_x M_{\sigma c},
\end{align}
which reproduces the Eq.(5a) in the main text.

\subsection{Double-cooperative process}
In the double-cooperative resolved case, the local fusion space
is the three-$\sigma$ fusion space with total charge fixed to $q=\sigma$, and we are left with a two-dimensional fusion space labeled by the intermediate fusion channel $x$ of the first two $\sigma$ anyons. The basis states can be chosen as
\begin{align}
\ket{((\sigma\sigma)_x\sigma)_\sigma},\quad x\in\{I,\psi\}. 
\end{align}

For the double-cooperative braid, the local two-exchange operator in this
basis are $B_{12}=R$ and $B_{23}=FRF^{-1}$. Therefore, for a resolved fusion channel $x$, the braid word $W=B_{12}B_{23}$ between the edge and QAD-bound anyons contributes the amplitude
\begin{align}
  \mathcal A_x^{\rm double}
  =
  \left[(FRF^{-1})R\right]_{xx},
\end{align}
Again, the central bulk charge $c$ is still not locally fused with the edge and QAD-bound anyons, so the full resolved-fusion interference amplitude is obtained by multiplying by an additional monodromy factor $M_{\sigma c}$, yielding
\begin{align}
  \mathcal I_x^{\rm double}
  =
  \left[(FRF^{-1})R\right]_{xx}M_{\sigma c}\label{seq:I_double_resolved},
\end{align}
which reproduces the Eq.(5b) in the main text.

\section{Decoherence effect: sequential regime}
\subsection{Sequentially resolved limit}

In addition to the resolved-fusion regime, Eq.~\eqref{seq:I_double_resolved} [Eq.(5b) in the main text] requires coherence between the three participating $\sigma$ anyons: while the edge anyon propagates from one QAD to the other, the propagating edge anyon and the two QAD-bound anyons form a coherent three-$\sigma$ fusion space. In
this regime, the total topological charge of the three $\sigma$ anyons is
fixed to be $\sigma$, and the second exchange acts on a different pair
inside the same three-anyon Hilbert space. This is why the $F$ move appears. This is natural for non-Abelian anyons: once a fusion channel is prepared or selected, the corresponding topological charge can be stored \textit{nonlocally}.

There is, however, another physically distinct limit where the coherence between anyons is lost due to some dephasing effects. If the fusion
coherence associated with the first local QAD event is lost or broken before the edge
anyon reaches the second QAD, the two local fusion resolutions should be
treated as two sequential, independent events rather than as a coherent
three-anyon evolution. We refer to this as the sequentially resolved limit.
In this limit, the first and second local fusion channels are labeled
independently by $x,y\in\{I,\psi\}$.
The first local exchange contributes $R^{\sigma\sigma}_{x}$, while the
second local exchange contributes $R^{\sigma\sigma}_{y}$. Therefore the
double-cooperative interference amplitude conditioned on the two local
fusion outcomes $(x,y)$ becomes a matrix
\begin{align}
  I^{\rm double,seq}_{xy}
  =
  R^{\sigma\sigma}_{x}R^{\sigma\sigma}_{y}M_{\sigma c}.
  \label{eq:seq_resolved_double}
\end{align}
This should be contrasted with the coherent result
$\bigl[(FRF^{-1})R\bigr]_{xx}M_{\sigma c}$, where braidings are performed coherently within the same
three-anyon fusion space. The results calculated from Eq.~\eqref{eq:seq_resolved_double} are summarized in
Table~\ref{tab:seq_resolved_double}.

\begin{table}[h]
\centering
\begin{tabular}{c|c|c|c}
\hline\hline
\diagbox[width=9em]{$(x,y)$}{Bulk charge} & $\quad c=I \quad $ & $\quad c=\psi\quad $ & $\quad c=\sigma\quad $\\
\hline
$(I,I)$
&
$e^{-i\pi/4}$
&
$-e^{-i\pi/4}$
&
$0$
\\
$(I,\psi)$ or $(\psi,I)$
&
$e^{i\pi/4}$
&
$-e^{i\pi/4}$
&
$0$
\\
$(\psi,\psi)$
&
$e^{3i\pi/4}$
&
$-e^{3i\pi/4}$
&
$0$
\\
\hline\hline
\end{tabular}
\caption{
Double-cooperative interference amplitudes in the sequentially resolved
regime, where the two local fusion channels $x$ and $y$ are resolved as
independent local outcomes.
}
\label{tab:seq_resolved_double}
\end{table}

If the local fusion outcomes $(x,y)$ are not postselected, the
measured signal is a classical average over these possible outcomes:
\begin{align}
  \overline I^{\rm double,seq}
  =
  M_{\sigma c}
  \sum_{x,y=I,\psi}p_{xy}
  R^{\sigma\sigma}_{x}R^{\sigma\sigma}_{y},
  \label{eq:seq_resolved_average}
\end{align}
where $p_{xy}$ is the probability of obtaining the local outcomes $(x,y)$.
If the two local events are statistically independent,
$p_{xy}=p_x^{(L)}p_y^{(R)}$, then we have 
\begin{align}
  \overline I^{\rm double,seq}
  =
  M_{\sigma c}
  \left[
  p_I^{(L)}R^{\sigma\sigma}_I
  +
  p_\psi^{(L)}R^{\sigma\sigma}_\psi
  \right]
  \left[
  p_I^{(R)}R^{\sigma\sigma}_I
  +
  p_\psi^{(R)}R^{\sigma\sigma}_\psi
  \right]\label{seq:I_probability}.
\end{align}
For equal probabilities ($p_I^{(L)}=p_I^{(R)}=p_\psi^{(L)}=p_\psi^{(R)}=1/2$), Eq.~\eqref{seq:I_probability} reduces to
\begin{align}
  \overline I^{\rm double,seq}
  =
  \frac{1}{2}e^{i\pi/4}M_{\sigma c},
\end{align}
which gives twice the exchange phase between Ising anyons but with an interference amplitude reduced by a factor of $1/2$.

\subsection{Sequentially unresolved limit}
For the unresolved-fusion regime, we find that a coherent
three-anyon evolution actually yields the same result obtained from two sequentially unresolved local events. 

When the two local unresolved fusion events are treated as
sequential and independent due to the broken coherence during anyon propagation, each contributes a quantum trace of elementary exchange $R$ in the two-$\sigma$ fusion space. Since the total charge of the two $\sigma$ anyons is $\sigma$, the quantum trace reduces to the ordinary trace. Hence, each local exchange contributes a twist factor
\begin{align}
  \frac{{\rm tr}R}{d_\sigma}
  =
  \frac{r_I+r_\psi}{\sqrt{2}}
  =
  e^{i\pi/8}
  =
  \theta_\sigma.
\end{align}
Thus, the sequentially unresolved local events gives
\begin{align}
  \left(\frac{{\rm tr}R}{d_\sigma}\right)^2
  =
  \theta_\sigma^2
  =
  e^{i\pi/4},
\end{align}
Including the bulk
monodromy, we have $\theta_\sigma^2M_{\sigma c}$, which is \textbf{identical} to the coherent unresolved result for the double-cooperative process [Eq.(4) in the main text].

This equivalence is physically understandable from the diagrammatic point
of view. In the unresolved-fusion regime, the intermediate fusion channel
is not a physical record; it is only an internal label introduced when one
chooses a particular fusion tree to evaluate the closed world-line diagram.
The double-cooperative process is therefore a closed $\sigma$ ribbon with
two positive twists, together with the same Hopf link with the bulk charge
$c$. How this closed ribbon is cut into intermediate fusion bases, or
whether one describes the two local events coherently or sequentially, does
not change the corresponding topological trace.

Mathematically, in the coherent three-anyon description, the $F$ move only
changes the fusion basis. The local contribution is
\begin{align}
  {\rm tr}\!\left[(FRF^{-1})R\right]
  =
  \sum_{x,y=I,\psi} r_x r_y |F_{xy}|^2 ,
\end{align}
where $r_x=R^{\sigma\sigma}_x$. For Ising anyons,
$|F_{xy}|^2=1/2$, and hence
\begin{align}
  {\rm tr}\!\left[(FRF^{-1})R\right]
  =
  \frac12(r_I+r_\psi)^2
  =
  \left(\frac{r_I+r_\psi}{d_\sigma}\right)^2
  =
  \theta_\sigma^2 .
\end{align}
This is exactly the result obtained by regarding the two local unresolved
events as two independent normalized quantum traces, each producing the
elementary twist factor
\begin{align}
  \frac{{\rm tr}R}{d_\sigma}=\theta_\sigma .
\end{align}
Thus the coherent and sequential unresolved descriptions give the same
universal twist factor $\theta_\sigma^2$.

The reason is not that the intermediate fusion channel is physically
irrelevant in general, but that in the unresolved regime no which-channel
information is measured, postselected, or retained. The fusion-channel
degree of freedom is fully traced out, so the basis dependence associated
with different fusion trees is removed by the quantum trace. This should be
contrasted with the resolved-fusion regime, where conditioning on a definite
local fusion channel inserts a physical projector onto $I$ or $\psi$ and
therefore retains which-channel information. In that case, coherent
three-anyon evolution and sequentially resolved local events are physically
distinct and generally lead to different amplitudes.

\section{Other platforms: topological superconductor and chiral spin liquid}

Another important realization of Ising anyons is provided by two-dimensional chiral topological superconductors that support Majorana zero modes (MZMs) bound to vortices or engineered defects~\cite{Ivanov_PRL_2001,ReadGreen2000,Alicea_2012}. The role of a QAD is most naturally played by a localized vortex or defect that traps a MZM. In this case, the Ising anyon does not carry an additional $U(1)$ sector, so the interferometric amplitude $\mathcal{I}$ is determined entirely by the non-Abelian Ising sector. Nevertheless, the main difficulties are the physical implementation of an ``edge anyon''~\cite{edge_anyon} and the choice of an appropriate observable. In particular, recent proposals suggest that mobile edge vortices --- also referred to as ``flying'' Ising anyons --- may provide a natural analogue of the propagating $\sigma$ anyon in superconductors~\cite{Andrea_PRL_2024,Adagideli2020,Beenakker_PRL_2019}. Possible readout mechanisms include charge sensing~\cite{vanLoo2026SingleShotKitaev}, parity-to-charge conversion~\cite{Munk2020}, or Josephson-type measurements~\cite{Demler_PRL_2013}.

On the other hand, the natural charge-neutral realization of Ising anyons can be found in the gapped chiral spin-liquid phase of the Kitaev honeycomb model~\cite{Kitaev_2006,banerjee2016proximate,chaloupka2010kitaev,kitagawa2018spin}, whose quasiparticles carry no electric charge but obey the same Ising fusion and braiding rules used above. In this case, the interferometric response should be interpreted not as an electrical conductance, but as a heat current $J_Q$, or equivalently the longitudinal thermal conductance $G_{xx}\equiv \partial J_Q/\partial(\Delta T)$ driven by a temperature bias~\cite{Klocke2021,Wei2021}. $\sigma$ anyons may in principle be trapped and manipulated by local defects, holes, or suitably engineered local magnetic-field textures~\cite{Wang2023}, thereby playing the role of quantum antidots. However, the gapless boundary mode corresponds to chiral Majorana fermions $\psi$ rather than Ising anyons $\sigma$. Such gapless chiral $\psi$ also exists in FQH systems, but it does not contribute to the electrical conductance due to its charge-neutral nature. The boundary of the FQH state supports the excitation of $\sigma$ whereas Kitaev spin liquid has a different boundary theory that only supports $\psi$. Therefore, implementing the interferometer setup in a spin-liquid platform would require additional ingredients such as phonon-coupling effect~\cite{Klocke2021}, $\sigma$-mediated tunneling~\cite{Wei2021,Wei_PRB_2023}.

\end{document}